\documentclass[12pt]{article}
\usepackage{amssymb}
\usepackage{enumerate}
\usepackage{dcolumn}% Align table columns on decimal point

\usepackage{graphicx}% Include figure files
\usepackage{dcolumn}% Align table columns on decimal point
\usepackage{bm}% bold math

\begin{document}

%\twocolumn

\def\H{{\cal H}}
\def\ttheta{\tilde{\theta}}

\def\beq{\begin{equation}}
\def\eeq{\end{equation}}
\def\bea{\begin{eqnarray}}
\def\eea{\end{eqnarray}}
\def\ben{\begin{enumerate}}
\def\een{\end{enumerate}}
\def\la{\langle}
\def\ra{\rangle}
\def\a{\alpha}
\def\b{\beta}
\def\g{\gamma}\def\G{\Gamma}
\def\d{\delta}
\def\e{\epsilon}
\def\phi{\varphi}
\def\k{\kappa}
\def\l{\lambda}
\def\m{\mu}
\def\n{\nu}
\def\o{\omega}
\def\p{\pi}
\def\r{\rho}
\def\s{\sigma}
\def\t{\tau}
\def\L{{\cal L}}
\def\S{\Sigma }
\def\gsim{\; \raisebox{-.8ex}{$\stackrel{\textstyle >}{\sim}$}\;}
\def\lsim{\; \raisebox{-.8ex}{$\stackrel{\textstyle <}{\sim}$}\;}
\def\gtrsim{\gsim}
\def\lessim{\lsim}
\def\loc{{\rm local}}
\def\vm{v_{\rm max}}
\def\bh{\bar{h}}
\def\del{\partial}
\def\nab{\nabla}
\def\half{{\textstyle{\frac{1}{2}}}}
\def\fourth{{\textstyle{\frac{1}{4}}}}

\begin{center} {\Large \bf
Neutron stars in Einstein-aether theory}
\end{center}

\vskip 5mm
\begin{center} \large
%\author
{{Christopher Eling$^{*}$\footnote{E-mail: cteling@physics.umd.edu},
Ted Jacobson$^{*}$\footnote{E-mail: jacobson@umd.edu}, and M.
Coleman Miller$^{\dagger}$\footnote{E-mail:
miller@astro.umd.edu}}}\end{center}

\vskip  0.5 cm
{\centerline{$^{*}$\it Department of Physics}}
{\centerline{\it University of Maryland}}
{\centerline{\it College Park, MD 20742-4111, USA}}
\vskip  0.5 cm
{\centerline{$^{\dagger}$\it Department of Astronomy}}
{\centerline{\it University of Maryland}}
{\centerline{\it College Park, MD 20742-4111, USA}}

\vskip 1cm

\begin{abstract}
As current and future experiments probe strong gravitational regimes
around neutron stars and black holes, it is desirable to have
theoretically sound alternatives to General Relativity against which
to test observations. Here we study the consequences of one such
generalization, Einstein-aether theory, for the properties of
non-rotating neutron stars. This theory has a parameter range that
satisfies all current weak-field tests. We find that within this
range it leads to lower maximum neutron star masses, as well as
larger surface redshifts at a particular mass, for a given nuclear
equation of state. For non-rotating black holes and neutron stars,
the innermost stable circular orbit is only slightly modified in
this theory.
\end{abstract}

\newpage

\section{Introduction}

General Relativity (GR) has passed every test so far, from solar
system dynamics and light bending to precision measurements of the
orbits of binary pulsars \cite{Kramer06}. Nonetheless, tests in
strong gravity remain elusive, largely because phenomena in strong
gravity involve other uncertain physics as well. For example,
spectral profiles of Fe~K$\alpha$ fluorescence lines in active
galactic nuclei and stellar-mass black holes
\cite{Iwasawa96,Fab02,RN03,RBG05,BR06} are consistent with the
expectations of gas streamlines near rapidly rotating black holes.
However, precision tests are not yet possible because of unknowns
about the emission profile and other complications.

More robust tests are on the horizon.  Larger-area X-ray detectors
such as {\it Constellation-X} \cite{White06} may be able to track
the motion of individual emitting elements in a disk, hence
mapping out the spacetime near a rotating black hole.  Even more
robustness is likely to come from detections of gravitational
waves from double black hole mergers of various masses and mass
ratios seen with ground-based interferometers such as LIGO
\cite{Barish00}, VIRGO \cite{Fidecaro97}, GEO-600
\cite{Schilling98}, and TAMA \cite{Ando02} and later with
space-based instruments such as LISA \cite{Danzmann00}. A third
source of strong field tests could exploit neutron star masses and
surface redshifts. It is that sort of system that is the subject
of this paper.

It is therefore desirable for current and future data to have
theoretically sound---and preferably well-motivated---alternatives
to or generalizations of GR. The simplest alternatives that have
been considered are various scalar-tensor theories. In the case of
the well-known Jordan-Brans-Dicke theory it has been shown
\cite{scalartensor} that the predictions of the theory in both the
weak and strong field regimes deviate from GR by a parameter that is
tightly constrained by post-Newtonian Solar System experiments.
Thus, the properties of compact objects such as neutron stars in
Jordan-Brans-Dicke theories that pass these weak field tests must be
very close to those found in GR. However, Damour and Esposito-Farese
\cite{Damour:1993hw} found a wide class of other scalar-tensor
theories exhibiting ``spontaneous scalarization," where weak field
constraints are met, but compact objects have significant deviations
from GR in the strong field regime. Recently, \cite{DeDeo:2003ju}
studied the properties of non-rotating neutron stars in these
theories, finding larger stellar masses than in GR and larger
surface redshifts for a given equation of state. These results were
then used to put an observational constraint on one of the
parameters of the model.

Over the past several years, a number of further alternatives to GR
have been proposed in the context of incorporating Lorentz violation
(LV) into gravity; see for example Refs.~\cite{Clayton:1999zs,
Jacobson:2000xp, Arkani-Hamed:2003uy, Gripaios:2004ms,Bluhm:2004ep,
Heinicke:2005bp,Rubakov:2006pn,Cheng:2006us} and references therein.
These theories can be thought of as low energy effective field
theory descriptions of local Lorentz violation (LV) possibly arising
from quantum gravitational physics at energies near the Planck
scale.  In this paper we study one of these models,
``Einstein-aether" theory (or ``ae-theory" for short)
\cite{Eling:2004dk}, in which a dynamical unit timelike vector field
$u^a$ is coupled to gravity. The LV here is a sort of spontaneous
symmetry breaking. The action is Lorentz invariant, but in any
solution defines the 4-velocity of a local ``preferred" frame at
each spacetime point, which breaks local boost symmetry.

The general ae-theory dynamics is dependent on four dimensionless
coupling parameters. Weak field constraints on ae-theory can all
be met by restricting these parameters to a subset of the four
dimensional parameter space that is very narrow compared to unity
in all but one dimension. The nature of these constraints is
summarized in the following section. They are derived in the weak
field regime, except for strongly self-gravitating neutron star
binaries. It is therefore interesting to ask what are the
deviations from general relativity in the strong field regime, and
to compare that behavior with observations of astrophysical
systems.

The main objective of this paper is to study the properties of
non-rotating neutron stars in ae-theory. The stellar solutions
depend on only one combination of the theory's coupling parameters.
We consider six candidate equations of state, three with purely
nucleonic degrees of freedom and different hardness, and three
involving quark matter with different bag constants. By numerical
solution of the field equations interior to the star we obtain the
maximum mass, relation between mass and radius, and surface
redshifts, all as a function of the coupling parameter.

It turns out that the maximum neutron star mass is less than in the
case of GR, and is smaller for larger values of the coupling
parameter. Thus, once the equation of state becomes known well
enough, it will be possible to place an upper bound on the coupling
parameter by observations of neutron star masses. Nonstandard
relations between mass and surface redshift also occur, providing
another possibility for  interesting phenomenology and constraining
the coupling parameter.

We also examine the location of the innermost stable circular
orbit (ISCO) as a function of mass, to determine whether that
might provide further useful observables distinguishing GR and
ae-theory. We find however that the ISCO is nearly unchanged for
reasonable coupling parameters and non-spinning objects.

The structure of this paper is as follows. In Section II we first
review the basics of Einstein-aether theory and summarize the
current constraints on theory. In Section III the equations of
structure for fluid stars in ae-theory are presented, together
with the form of the analytic exterior solution to which the
interior must be matched. Using the exterior solution, expressions
are obtained for the surface redshift and ISCO radius that can be
employed with the numerical solutions to obtain the observable
quantities. We present the numerical results in Section IV,
together with the constraints on the coupling parameter that can
be obtained with these results. In Section V we conclude with a
brief discussion of prospects for further constraints from more
precise neutron star measurements and rotating black hole
solutions.

\section{Einstein-aether theory}
\label{aetheory}

The action for Einstein-aether theory is the most general
generally covariant functional of the spacetime metric $g_{ab}$
and aether field $u^a$ involving no more than two derivatives (not
including total derivatives),
\beq S = \frac{1}{16\pi G}\int \sqrt{-g}~ (L_{\ae}+L_{\rm matter})
~d^{4}x \label{action} \eeq
where
\beq L_{\rm ae} = -R-K^{abmn} \nabla_a u_m \nabla_b u_n -
\lambda(g^{ab}u_a u_b - 1) \eeq
and $L_{\rm matter}$ denoted the matter lagrangian. Here $R$ is
the Ricci scalar, $K^{ab}_{mn}$ is defined as
\beq K^{ab}_{mn} = c_1 g^{ab}g_{mn}+c_2\d_m^a\d_n^b
+c_3\d_n^a\d_m^b +c_4u^au^bg_{mn} \eeq
where the $c_i$ are dimensionless coupling constants, and
$\lambda$ is a Lagrange multiplier enforcing the unit timelike
constraint on the aether. The convention used here for metric
signature is $({+}{-}{-}{-})$ and the units are chosen so that the
speed of light defined by the metric $g_{ab}$ is unity. Note that
since the covariant derivative operator $\nabla_a$ involves
derivatives of the metric through the connection components, and
since the unit vector is nowhere vanishing, the terms quadratic in
$\nabla u$ also modify the kinetic terms for the metric.

The matter Lagrangian $L_{\rm matter}$ generically will be a
functional of a collection of matter fields (denoted as $\psi$)
along with $g_{ab}$ and $u^a$. However, following the observational
constraints on Lorentz violation in the matter sector, we assume
here when studying the neutron star solutions that there is no
significant coupling of matter to $u^a$. The absence of coupling of
$u^a$ to matter has no theoretical justification in this purely
phenomenological approach, and may be regarded as unnatural. However
our goal here is just to explore consequences of gravitational
Lorentz violation in a phenomenologically viable setting. It remains
an open question whether this can emerge as an approximation to a
more fundamental underlying theory.

The field equations from varying (\ref{action}) with respect to
$g^{ab}$, $u^a$ and $\lambda$ are given by
\begin{eqnarray}
G_{ab} &=& T^{(u)}_{ab}+8\pi G T^{M}{}_{ab}\label{AEE}\\
\nab_a J^{a}{}_m-c_4 \dot{u}_a \nab_m u^a &=& \l u_m,
\label{ueqn}\\
g_{ab} u^a u^b &=& 1, \label{constraint}
\end{eqnarray}
where
\beq J^a{}_{m} = K^{ab}{}_{mn} \nabla_b u^n. \label{Jdef}\eeq
The aether stress tensor is given by
\bea T^{(u)}{}_{ab}&=&\nab_m(J_{(a}{}^m u_{b)}-J^m{}_{(a} u_{b)}
- J_{(ab)}u^m) \nonumber\\ &&+ c_1\, \left[(\nab_m u_a)(\nab^m u_b) - (\nab_a u_m)(\nab_b
u^m) \right]\nonumber\\ &&+ c_4\, \dot{u}_a\dot{u}_b\nonumber\\
&&+\left[u_n(\nab_m J^{mn})-c_4\dot{u}^2\right]u_a u_b \nonumber\\
&&-\frac{1}{2} L_u g_{ab}, \label{aetherT}\eea
where $L_{u} = -K^{ab}{}_{mn} \nabla_a u^m \nabla_b u^n$. The
Lagrange multiplier $\lambda$ has been eliminated from
(\ref{aetherT}) by solving for it via the contraction of the
aether field equation (\ref{ueqn}) with $u^a$.

\subsection{Observational constraints on the parameters $c_i$}
\label{constraints}

In the weak-field, slow-motion limit ae-theory reduces to
Newtonian gravity~\cite{Carroll:2004ai}, with a value of  Newton's
constant $G_{\rm N}$ related to the parameter $G$ in the action
(\ref{action})  by
\beq G_{\rm N}=\frac{G}{1-(c_1+c_4)/2}. \label{GN}\eeq
The phenomenology of Einstein-aether theory has been extensively
studied over the last few years.  Theoretical and observational
constraints on the coupling parameters $c_i$ have been determined
from parameterized post-Newtonian analysis~\cite{Eling:2003rd,
Graesser:2005bg,Foster:2005dk}, stability and linearized energy
positivity~\cite{Jacobson:2004ts,Lim:2004js, Elliott:2005va,
Eling:2005zq}, primordial nucleosynthesis~\cite{Carroll:2004ai},
and vacuum Cerenkov radiation~\cite{Elliott:2005va}. The combined
constraints from all of these are reviewed in
Ref.~\cite{Foster:2005dk}.

To summarize here, all parameterized post-Newtonian (PPN) parameters
except the preferred frame parameters $\alpha_{1,2}$ agree with
those of GR for any choice of the $c_i$. Observations impose strong
constraints on $\a_1$ ($\lesssim 10^{-4}$) and $\a_2$ ($\lesssim
4\times 10^{-7}$). These parameters can be set to zero in
Einstein-aether theory by imposing two conditions, on the $c_i$,
which can be solved to determine
\begin{eqnarray}
 c_2&=&(-2c_1^2-c_1c_3 + c_3^2)/3c_1 \nonumber\\
 c_4&=&-c_3^2/c_1 \label{zeroalphas}.
\end{eqnarray}
With this choice the gravitational constant
appearing in the cosmological Friedman equations agrees
with that appearing in the force law between
isolated masses, so there is no further nucleosynthesis
constraint. The stability, positive energy, and vacuum
\v{C}erenkov constraints then impose the inequalities
\begin{eqnarray}
0&<&c_{+}<1 \nonumber\\
0&<&c_{-}< c_{+}/3(1-c_{+}), \label{superluminal}
\end{eqnarray}
where $c_\pm=c_1\pm c_3$.

\begin{figure}
  % Requires \usepackage{graphicx}
\begin{center}
 \includegraphics[angle=-90,width=13cm]{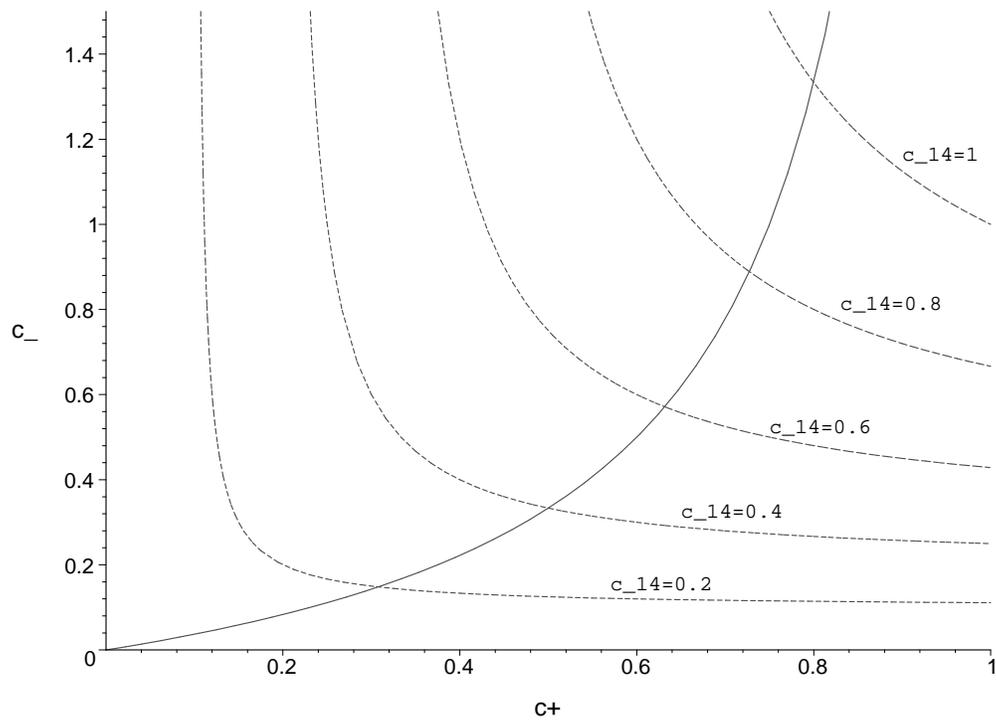}\\
\end{center}
\caption{\label{fig:constraint} Graphical representation of
(\ref{superluminal}) (solid line) and (\ref{c14}) (dashed lines).
The allowed region of the parameter space is below the solid curve,
above $c_-=0$ and to the left of $c_{+}=1$. The radiation damping
constraint will restrict to a nearly one-dimensional subset of this
region.}
\end{figure}

Further constraints have been obtained using radiation damping in
binary pulsar systems~\cite{Foster:2006az}. An analysis neglecting
strong self-gravitating effects found that when (\ref{zeroalphas})
hold, just one condition ${\cal A}(c_1,c_3)=1$ makes the lowest
order radiation rate in ae-theory identical to that of GR. This
condition is satisfied entirely in the region allowed by
(\ref{superluminal}).\footnote{In  \cite{Foster:2006az}, the ${\cal
A}=1$ curve does not fall entirely in the otherwise allowed region,
but this is due to an error in the analysis there that has since
been corrected~\cite{Foster:2007gr}.} However, the neutron star
sources are strongly self-gravitating. It turns
out~\cite{Foster:2007gr} that as long as $c_i\lessim 0.1$, the
strong field corrections are negligible, but for larger coupling
values the precise radiation damping constraints are not yet worked
out.  They will lead to a modified condition ${\cal A}'(c_1,c_3)=1$
that will depend on the nature of the compact objects in the binary.

Neutron star structure constraints along the lines discussed in
the present paper should eventually be able to restrict
$c_{14}=c_1+c_4$, which is given by
\beq
c_{14}=2c_+c_-/(c_++c_-)
\label{c14}
\eeq
when the PPN equivalence conditions (\ref{zeroalphas}) hold. The
only previous constraint on $c_{14}$ was the requirement that it
be less than $2$ in order to maintain positivity of Newton's
constant (\ref{GN}).

Fig.~\ref{fig:constraint} shows the region in the $(c_{+},c_{-})$
parameter space allowed by the above constraints (other than the
radiation damping constraint), along with $c_{14}$ contours. Note
that without a constraint on $c_{14}$,  $c_{-}$ can grow arbitrarily
large as $c_{+} \rightarrow 1$. Any upper bound on $c_{14}$ will cut
off this region however.  A $c_{14}$ contour intersects the right
hand boundary ($c_+=1$) of the allowed region at
$c_-=c_{14}/(2-c_{14})$, and intersects the upper boundary at
$c_-=(4/3)c_{14}/(2-c_{14})$.

\section{Neutron Stars}

Time independent spherically symmetric solutions in ae-theory were
extensively studied recently in a pair of papers on fluid
star~\cite{Eling:2006df} and black hole~\cite{Eling:2006ec}
configurations. For black holes, the aether vector has a radial
component, but for the case of a star it does not, i.e.  it is
aligned with the static frame defined by the Killing vector, both
inside and outside the star. Thus, unlike in GR, the exterior
solution for a star is not the same as for a black hole.

In Ref.~\cite{Eling:2006df} the static vacuum exterior solution
was found analytically, and numerical integration was employed to
find interior stellar solutions for the test case of a constant
mass density equation of state. The mass was found by matching the
interior to the exterior solutions. Unlike constant density stars
in General Relativity (GR), the mass does not increase
monotonically with central pressure; rather, there is a maximum
mass at a finite central pressure beyond which the stars are
unstable. This maximum mass is smaller than in the GR case.

It is adequate for our present purposes to restrict attention to
non-rotating stars, since the effect of rotation on the maximum
mass, surface redshift, and ISCO is very small for the
observationally relevant spins. For example, presuming the
fractional change in maximum mass scales with the square of the
spin, Tables 4 and 5 in Ref.~\cite{Cook:1993qr} indicate that even
for a one millisecond period the increase of maximum mass is less
than 5\% in GR. Barring an unexpected much greater sensitivity to
a small spin in ae-theory, the results we find here for
non-rotating stars should be quite reliable except for the fastest
spinning stars.

\subsection{Stellar Equations of Structure}

The static, spherically symmetric form for the metric and aether
can be written in Schwarzschild-like coordinates as
\bea ds^2 &=& e^{A(r)} dt^2 - B(r) dr^2 - r^2 d\Omega^2,\label{metricansatz}\\
u &=& e^{-A(r)/2}~\partial_t. \label{aetheransatz} \eea
Note that since the aether is aligned with the timelike Killing
vector it is completely determined by the metric. In particular,
the aether is at rest with respect to the static frame at
infinity, which means that the star is taken to be at rest with
respect to the aether. When comparing theory and observation, it
is typically assumed that the background aether frame coincides
with that of the cosmological fluid. Any particular star will of
course have some proper motion with respect to this frame, so
strictly speaking the physically relevant solutions are not of the
form (\ref{metricansatz},\ref{aetheransatz}). However, assuming a
relative velocity of order $10^{-3}$, this discrepancy should not
be significant for comparisons of much less precision such as
concern us here.

It was shown in \cite{Eling:2006df} that for configurations of the
form  (\ref{metricansatz},\ref{aetheransatz}) the $c_2$ and $c_3$
terms in the action (\ref{action}) and their variations are zero,
and thus they do not contribute to the field equations. Also, the
effect of the $c_4$ term can be absorbed by the replacement
$c_1\rightarrow c_1+c_4$. Hence the only coupling relevant to
these solutions is $c_{14}\equiv c_1+c_4$. The fluid stress tensor
appearing in the metric field equation (\ref{AEE}) is
\beq T^M_{ab} = \bigl(\rho(r)+P(r)\bigr) v_a v_b - P(r)
g_{ab}\label{perffluid}\eeq
where $v^a = e^{-A/2}(\partial_t)^a$ is the fluid 4-velocity,
$\rho(r)$ its mass density, and $P(r)$ its pressure.

The metric field equation and the Bianchi identity together imply
that the sum of the aether and fluid energy-momentum tensors is
divergenceless. In addition, since the aether does not couple
directly to the fluid, its stress tensor is independently
divergenceless when its field equation and unit constraint are
satisfied. Therefore the fluid stress tensor is also independently
divergenceless in any solution. Thus, an appropriate system of
equations for the aether plus fluid case is the (i) metric field
equation, (ii) aether field equation, (iii) radial component of
$\nabla^a T^M_{ab}=0$, which is the hydrostatic equilibrium
equation for the fluid
\beq P'+\half A'(\rho+P) = 0, \label{hydro} \eeq
and (iv) an equation of state $\rho=\rho(P)$.

It turns out that the aether field equation (\ref{ueqn}) has only
a $t$ component, which just determines the Lagrange multiplier
$\lambda$. The $tt$, $rr$, and $\theta\theta$ components of the
metric field equation then imply respectively
\begin{eqnarray}
 0&=&-1+B+r\frac{B'}{B} -\n(8 r A' + r^2 A'^2 -2r^2
A' \frac{B'}{B}+4 r^2 A'') -  \rho r^2 B\label{tt}\\
0&=& 1-B+ rA' +\n r^2 A'^2-  P r^2B \label{rr}\\
0&=&  \frac{1}{4}(2rA'-2r\frac{B'}{B}+ r^2 A'^2 -r^2
A'\frac{B'}{B} +2r^2A'') -\n r^2 A'^2   -  PB,\label{thetatheta}
\end{eqnarray}
where the symbol
\beq \n=\frac{c_{14}}{8} \eeq
is introduced to compactify the notation, and we have adopted
units with $8\pi G=1$.

Equation (\ref{rr}) can be used to solve for $B$,
\beq B  = (1+ r^2 P)^{-1}(1+rA'+\n r^2A'^2 ). \label{Beqnin}\eeq
After substituting this result into (\ref{tt}) and
(\ref{thetatheta}), $A''$ can be eliminated from this pair of
equations, yielding an equation involving $A$, $A'$,  $P$,  $P'$,
and $\rho(P)$. Due to its complexity it does not seem illuminating
to display it here. This equation combined with (\ref{hydro}) can
then be numerically integrated to solve for $P(r)$ and $A(r)$
starting with initial values at the origin $r=0$. (It is possible
to eliminate $A'$, leaving one Tolman-Oppenheimer-Volkoff (TOV)
type equation for $P(r)$. Since this TOV equation is quite
complicated and doesn't aid the numerical integration procedure we
will not display it here.)

To numerically integrate outward we find the power series solution
to the equations (\ref{hydro}) and
(\ref{tt},\ref{rr},\ref{thetatheta}) in the vicinity of $r=0$,
which is a singular point for the equations. In this solution the
central value for the pressure $P(0)=P_0$ is the only free
parameter to be specified (A(0) is arbitrary due to scaling
freedom of the $t$ coordinate, so can just be set to unity). The
numerical integration can then be started at a small value of $r$
using the power series for initial data, and continued to the
value $r=R$, which is the surface of the star where the pressure
and mass density drop to zero. At this point the values of $A$ and
$A'>0$ can be matched to the static vacuum aether solution
discussed in \cite{Eling:2006df} and summarized in the next
section.

\subsection{Vacuum solution}

In the exterior (\ref{Beqnin}) becomes
\beq B = 1+rA'+ \n\,  r^2A'^2, \label{Beqnout}\eeq
and the remaining field equations (\ref{tt}) and (\ref{thetatheta})
can be reduced to the second order ordinary differential equation
(ODE)
\beq r^2A'' +  2rA' + r^2 A'^2 +   \n\,  r^3 A'^3=0,
\label{staticeom}\eeq
With the substitution
\beq Y=rA', \label{A'} \eeq
(\ref{staticeom}) becomes a first order equation for $Y(r)$ that
can solved, after which (\ref{A'}) can solved for $A(Y)$. In terms
of the roots of $B$,
\beq Y_\pm=(-1\pm\sqrt{1-4\n})/(2\n), \eeq
the result is
 \beq B=\n(Y-Y_-)(Y-Y_+),\eeq
\beq
N=e^A=\left(\frac{1-Y/Y_-}{1-Y/Y_+}\right)^{\frac{-Y_+}{2+Y_+}},
\label{N} \eeq
and
\beq \frac{r_{\rm
min}}{r}=\left(\frac{Y}{Y-Y_-}\right)\left(\frac{Y-Y_-}{Y-Y_+}\right)^{\frac{1}{2+Y_+}},
\label{C/r} \eeq
where $r_{\rm min}$ is an integration constant. Thus the complete
solution is known up to the inversion of the function on the right
hand side of (\ref{C/r}).

As in GR, the solutions in this family are all asymptotically
flat. The limit $Y\rightarrow 0$ corresponds to spatial infinity,
where the limiting form of the solution is
\bea
B&=&1+Y +\cdots\\
N&=&1-Y+\cdots\\
Y&=&r_g/r+\cdots. \eea
Here $r_g$ is the gravitational radius, which is related to the
gravitational mass $M$ appearing in the Newtonian potential by
\beq r_g=2G_{\rm N}M. \eeq
The relation between $r_g$ and the minimum radius\footnote{In the
pure vacuum solution, there is a minimal 2-sphere ``throat" at
$r=r_{min}$. In a fluid star solution the radius of the star is
always greater than $r_{min}$, so there is no throat in the spatial
geometry~\cite{Eling:2006df}.} $r_{min}$ is given by
\beq r_{\rm min}/r_g=(-Y_+)^{-1}(-1-Y_+)^{(1+Y_+)/(2+Y_+)}.
\label{M} \eeq
To leading order in $1/r$ this solution agrees with the
Schwarzschild solution of GR. The total gravitational mass $M$ of
the fluid star can be read off from (\ref{M}) together with
(\ref{C/r}), using the definition (\ref{A'}), $Y(R)=RA'(R)$.

\subsubsection{Surface redshift}
The light emitted from the surface of the neutron star is
redshifted as it climbs away to a distant observer. From
(\ref{metricansatz}), the surface redshift factor $z$ is given by
\beq z = [N(R)]^{-1/2}-1,\eeq
which can be evaluated directly from the numerical solution using
(\ref{N}) and (\ref{A'}).

\subsubsection{ISCO}
\label{ISCO}
 The orbits in the metric (\ref{metricansatz}) have
conserved energy $e=N \dot{t}$  and angular momentum
$\ell=r^2\dot{\phi}$, where $N=e^A$ and the overdot stands for
derivative with respect to proper time. Since the parameter is
proper time, the four-velocity has unit norm. This condition can
be expressed in the form
\beq \dot{r}^2=V(r)=B^{-1}W, \eeq
with
\beq W=W(r; e,\ell)=N^{-1}(r)e^2-r^{-2}\ell^2+1. \eeq The ISCO is
determined by the conditions $V=V'=V''=0$, or equivalently,
$W=W'=W''=0$, where the prime stands for derivative with respect
to $r$. Thus the metric function $B$ plays no role. These
equations determine $r$, $e$ and $\ell$ at the ISCO. After some
manipulation of the equations we obtain
\beq Y_{\rm ISCO}=\frac{-1+\sqrt{1+\nu}}{\nu} \eeq
With this result, the radius of the ISCO can be found from
(\ref{C/r}) and $(\ref{M})$ given the mass. Expanding in
$\nu=c_{14}/8$ we find (in units with $G_{\rm N}M=1$)
\beq r_{\rm ISCO}\simeq 6(1 + [\ln(3/2)-1/6] \nu)\simeq 6
(1+0.030\, c_{14}), \eeq
dropping $O(\nu^2)$ terms. This linear approximation is extremely
accurate: the relative error grows monotonically from 0 to only
about 0.3\% over the entire allowed range of $c_{14}$ from 0 to
$2$.

The angular frequency of an orbit with respect to time at infinity
is given by $\o=\dot{\phi}/\dot{t}=(\ell/e)(N/r^2)$. The circular
orbit condition yields $N\ell/e=(N'r^3/2)^{1/2}=r(YN/2)^{1/2}$, so
$\o= r^{-1}(YN/2)^{1/2}$. Expanding again in $\nu$, the frequency
at the ISCO is found (in units with $G_{\rm N}M=1$) to be
\beq \o_{\rm ISCO} \simeq\frac{1}{6\sqrt{6}}
(1+[-2\ln(3/2)+1/2]\nu) \simeq \frac{1}{6\sqrt{6}} (1-0.039\, c_{14}),
\label{f} \eeq
dropping $O(\nu^2)$ terms.

Thus, even for the maximum value $c_{14}=2$, the location of the
ISCO is only about 6\% larger than its value in GR for a star of
the same mass, and the orbital frequency is about 8\% smaller.
Since ae-theory agrees so closely with GR on these quantities, it
is unlikely that in the near future any useful constraints can be
obtained from their behavior for slowly rotating stars.  However,
as we discuss in Section \ref{conclusions}, for the ISCO of a
rapidly rotating black hole the deviations from GR may well be
much more pronounced.

\section{Numerical results}

Here we will compare the properties of neutron stars in GR and
ae-theory using three hadronic and three quark equations of state
(EOS). We label these according to whether they are softer (s),
medium (m), or harder (h), by
\bea {\rm Hs,Hm,Hh}&\leftrightarrow&
{\rm A18, A18}\d v{\rm UIX,A18UIX}\\
{\rm Qs,Qm,Qh}&\leftrightarrow& {\rm (90,0), (60,200), (60,0)}.
\eea
The hadronic models are discussed in~\cite{Akmal:1998cf}, and the
quark models are MIT bag models~\cite{Farhi:1984qu} determined by two
parameters $(B,m_s)$, with the bag constant $B$ measured in
MeV/fm$^3$ and the strange quark mass $m_s$ in MeV.

The pressure and mass density data tables for these
models~\cite{Bhattprivate} were converted from cgs units to
geometrized units, i.e. replacing the energy density $\rho$ and
pressure $p$ by $G_{\rm N}\rho/c^4$ and $G_{\rm N}p/c^4$
respectively, yielding quantities with dimension inverse length
squared. A curve fitting procedure was then used to generate an
equation of state function $\rho(P)$ suitable for the numerical
integration. The field equations (\ref{tt}-\ref{thetatheta}) are
written in units with $8\pi G=c=1$, so to apply them we first
multiply the density and pressure in the above geometrized units
by $8\pi G/G_{\rm N}=8\pi(1-c_{14}/2)$.

\begin{figure}
  % Requires \usepackage{graphicx}
 \includegraphics[angle=-90,width=13cm]{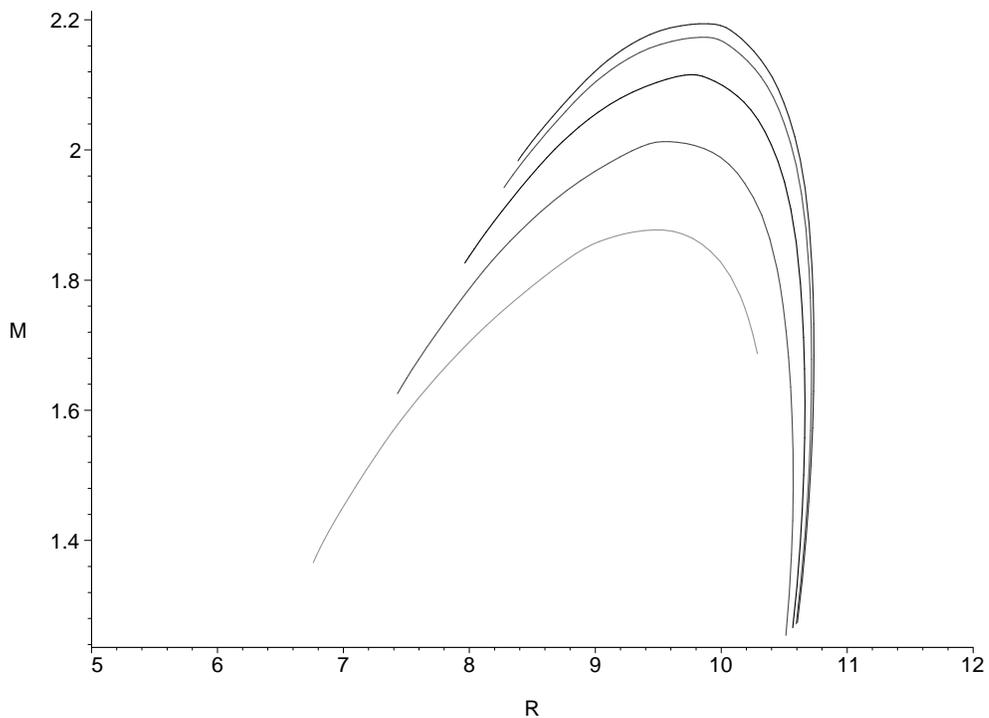}\\
\caption{\label{massvR} Total mass versus $R$ for the Hm equation of
state for $P_0$ up to 100 and $c_{14}=0,\, 0.05,\, 0.2,\, 0.5,\, 1$
beginning with the upper (GR) curve. The vertical axis is units of
solar masses and the horizontal in km. The GR curve reaches its
maximum mass of $2.20 M_\odot$ at slightly more than 10 km. As
$c_{14}$ increases to 1 the maximum mass decreases and the value of
the radius at the maximum falls to about 9.5 km.}
\end{figure}

In GR five of the six equations of state have associated $M$ versus
$P_0$ curves containing a maximum mass extremum and regions of
stability and instability.  The exception is the softest quark EOS,
Qs,  which appears to asymptote to its maximum mass value. The GR
maximum mass values for the Hm and Hs equations of state we find
here ($2.20 M_\odot$ and $1.67 M_\odot$ respectively) agree very
well with the results obtained in \cite{Akmal:1998cf}. The maximum
mass value in ae-theory grows smaller and occurs at smaller values
of central pressure as $c_{14}$ is increased. For the Qs EOS extrema
begin to develop in the $M$ versus $P_0$ curve as $c_{14}$
approaches 1. Fig. \ref{massvR} shows a plot of $M$ vs. $R$ for the
Hm EOS, each point being determined by a value of $P_0$, for several
values of $c_{14}$.
As $P_0$ increases the mass values increase sharply to peaks and
then gradually fall off. The region of the curves for small $P_0$
and larger $R$ up to the mass maximum describe stable equilibrium
configurations. Beyond the maximum the neutron stars are unstable.
In ae-theory the minimum radius where the equilibrium
configuration is stable decreases as $c_{14}$ increases.

\begin{figure}
  % Requires \usepackage{graphicx}
 \includegraphics[angle=-90,width=13cm]{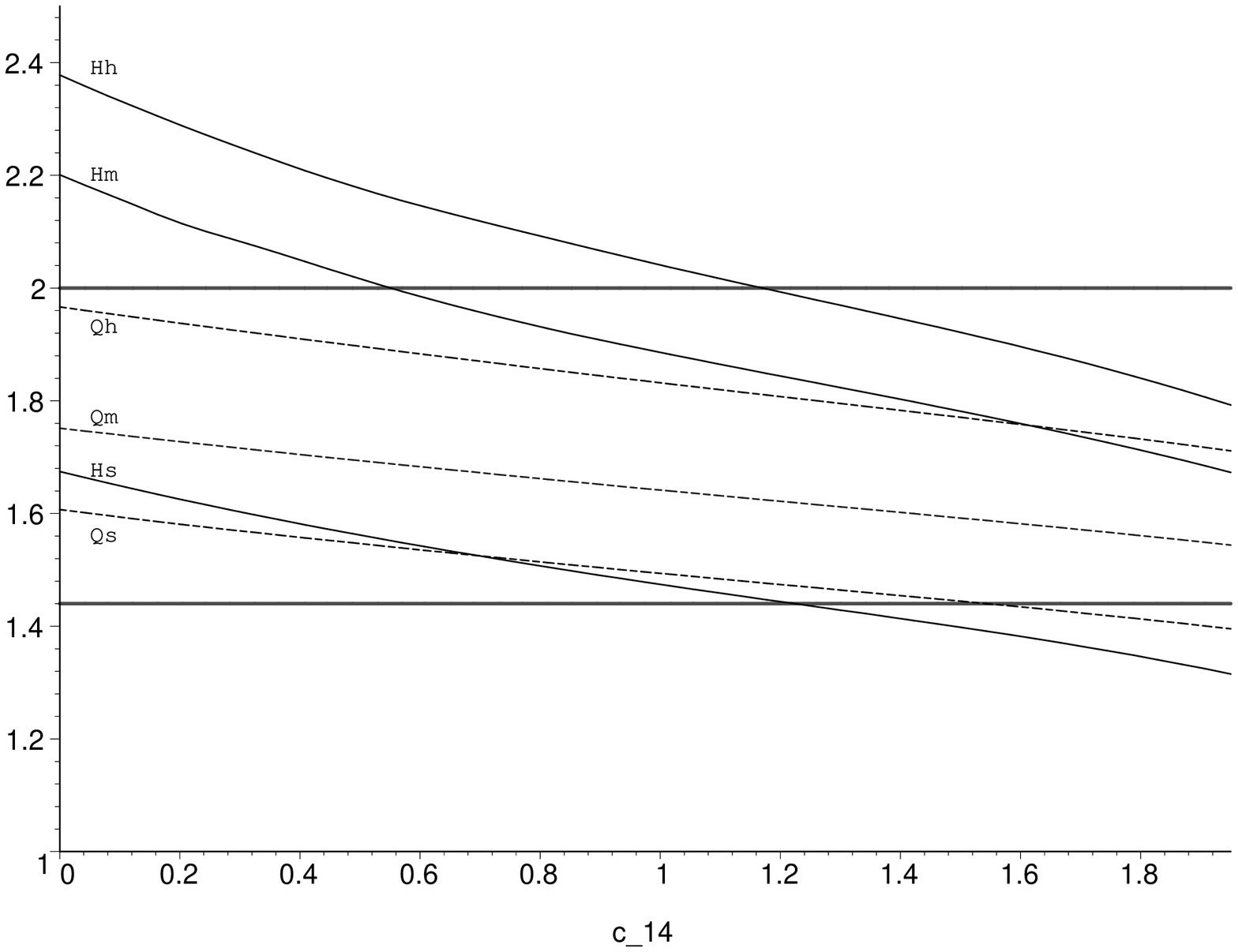}\\
\caption{\label{massconstraints} Maximum mass vs.\ $c_{14}$ for the
six equations of state. The hadronic models are plotted with solid
lines, while the quark models are dashed. The thick solid horizontal
lines represent the bare minimum constraint of $1.44\,M_\odot$ and a
possible constraint value of $2\,M_\odot$. }
\end{figure}

A plot of the maximum mass values for the six equations of state
considered in this paper is shown in Fig. \ref{massconstraints}.
Horizontal lines mark the certain lower bound of $1.44\,M_\odot$
and a benchmark value of $2.0\,M_\odot$. The dependence of the
maximum mass on $c_{14}$ is very close to linear for the quark
models, with the mass changing by roughly 6\% as $c_{14}$
increases from 0 to 1. For the hadronic models it is roughly
linear but steeper, decreasing by roughly 15\% over the same range
of $c_{14}$.

\subsection{Maximum mass constraints}

The most straightforward constraint on ae-theory comes from
comparing the maximum mass values generated with the six equations
of state to observations of neutron star masses in binary pulsars.
These masses are not directly measured, but inferred from the
timing data from a binary pulsar system. This data contains
information on the Keplerian and post-Keplerian parameters of the
system, which depend on the unknown masses $m_A$ and $m_B$ of the
neutron stars. A determination of the Keplerian and two
post-Keplerian parameters, such as the secular rate of periastron
advance and the magnitude of the Shapiro delay, result in two
curves in a $(m_A, m_B)$ mass plane. The value of the two masses
is the intersection of these curves. Since we are considering the
subset of Einstein-aether theories that satisfies the
post-Newtonian constraints, any corrections to how the
post-Keplerian parameters depend on the pulsar masses would only
appear at higher order and is therefore very small. Thus, the
masses can be inferred as in GR to a good approximation. If the
maximum mass predicted by Einstein-aether theory is smaller than
an observed neutron star mass then the theory is ruled out.
Currently the largest reliable observed mass value is
$1.44,M_\odot$ from the PSR 1913+16 Hulse-Taylor binary system.

There are, however, suggestions from various data that neutron
star masses can be at least $\sim 2\,M_\odot$.  The neutron star
in Vela~X-1 has an estimated mass of $1.88\pm 0.13\,M_\odot$
\cite{Quaintrell03}, and PSR~J0751, a pulsar in a detached
low-mass binary, has a reported mass of $2.1\pm 0.2\,M_\odot$
\cite{Nice05}. Furthermore, there are indications (although not as
definitive) for neutron star masses greater than or of order two
solar masses in several low-mass X-ray binaries based on the
inference of the orbital frequency at the ISCO from their
kilohertz quasi-periodic brightness oscillations
(QPO's)\cite{Barret05a,Barret05b,Barret06,Barret07}; for an
alternative view, see \cite{Mendez06}. At the $<700$~Hz spin
frequencies of these stars, the dimensionless angular momentum is
only 0.1 to 0.3 \cite{Cook:1993qr}, so this would imply that the
ISCO is obtained from a near-Schwarzschild spacetime. Since the
ISCO we find in Section \ref{ISCO} is very close to the GR value,
the derived mass should be the same in ae-theory as in GR to a
good approximation. As a benchmark, we will consider the limits on
$c_{14}$ that would result from a measured gravitational mass of
$2\,M_\odot$.

Fig.~\ref{massconstraints} shows that in GR ($c_{14} = 0$) all six
equations of state respect the lower bound of $1.44\,M_\odot$ solar
masses. For four of the equation of state models the $1.44\,M_\odot$
mass cutoff does not yield any constraint on $c_{14}$. For the Hs
and Qs EOS there are weak constraints that $c_{14}$ be less than
about 1.2 and 1.5 respectively. The $2\,M_\odot$ constraint is more
restrictive. In this case the Hs, Qs, Qm, and Qh EOS are ruled out,
while for  Hm and Hh $c_{14}$ must be less than about 0.55 and 1.16
respectively.

As the maximum observed neutron star mass is pushed upwards, and
more is learned about the nuclear EOS, the observational upper bound
on $c_{14}$ will come down. If we assume the existence of a
non-rotating neutron star of $2\,M_\odot$, then even for the hardest
equation of state we have considered we obtain the bound
$c_{14}<1.16$.

\subsection{Surface redshift constraints}

There is not yet a definitive detection of an atomic spectral line
from the surface of a neutron star.  The strongest current case
comes from stacked observations of thermonuclear X-ray bursts from
EXO~0748--676, from which a surface redshift of 0.35 was
inferred~\cite{Cottam02} based on identification of some
absorption-like features as being produced by highly ionized iron.
The mass of this star is not certain, but \"Ozel \cite{Ozel:2006km}
used simplifying assumptions about the constancy of the peak flux of
the bursts and radiative transfer to suggest that the mass of this
object is probably not less than $1.8\,M_\odot$.

\begin{figure}
  % Requires \usepackage{graphicx}
\includegraphics[angle=-90,width=12cm]{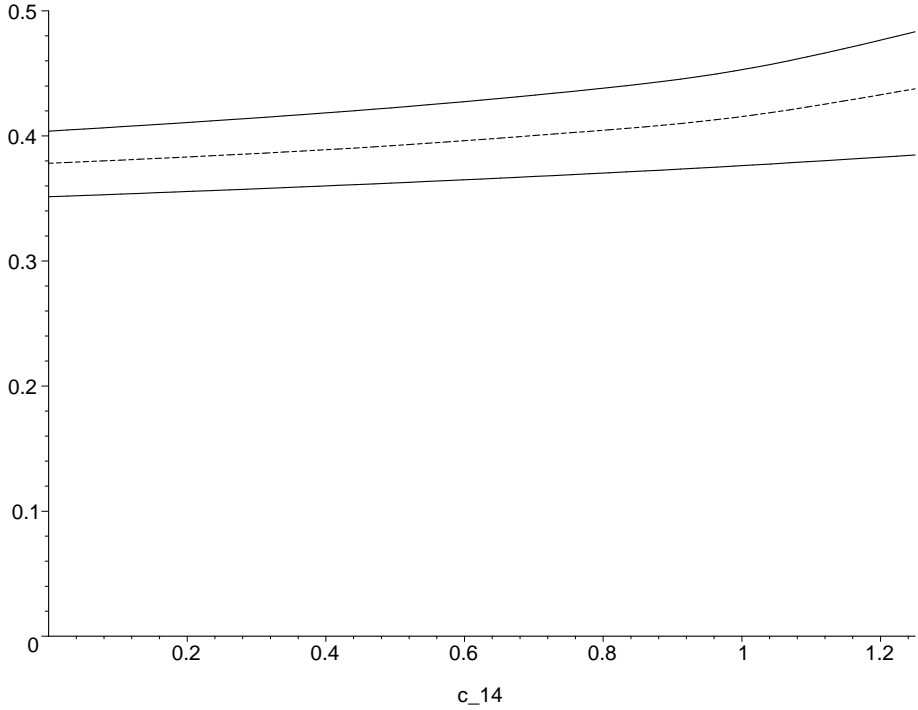}\\
\caption{\label{redshift}
Surface %t27
redshift factor $z$ versus $c_{14}$ for $1.8\,M_\odot$ neutron stars
using the hardest equations of state. Hm (solid) is on top, Qh
(dashed) is in the middle and Hh (solid) is on the bottom. Note that
the GR value of $z=0.35$ for the hardest EOS, Hh, is consistent with
the
%t27 observed
proposed surface redshift of 0.35 for EXO~0748--676 \cite{Cottam02}.
The Hm and Qh lines begin to curve up near $c_{14} = $1.1-1.2
because the maximum mass for these equations of state is approaching
$1.8\,M_\odot$.}
\end{figure}

Measurements such as these, once confirmed, can provide a joint
constraint on $c_{14}$ and the equation of state via the dependence
of surface redshift on mass in ae-theory. (We have checked that the
mass inferred using the method of Ref.~\cite{Ozel:2006km} differs
from the GR value by less than 2\% when $c_{14}=1$, so the leading
order effect of $c_{14}$  is only in the relation between radius and
mass, equivalently redshift and mass.) Fig. \ref{redshift} shows a
plot of $z$ versus $c_{14}$ for $1.8\,M_\odot$ stars using the Hm,
Hh, and Qh EOS. These are the three hardest equation of state models
and have equilibrium configurations at this mass. (As shown in
Fig.~\ref{massconstraints}, the other three softest equation of
state models do not have equilibrium configurations at this mass.)
The surface redshifts increase by roughly 10\% as $c_{14}$ ranges
from 0 to 1.

If in the future surface redshifts together with masses can reliably
be determined, then tight constraints on the equation of state in GR
may be obtained by combining measurements for a collection of stars.
It is also possible that single measurements may provide stringent
constraints. For example, as revealed in Fig.~\ref{redshift}, the
proposed surface redshift 0.35 of EXO~0748--676~\cite{Cottam02} is
compatible with $1.8\,M_\odot$ only for the hardest eos (Hh) among
those we considered. In general, the parameter $c_{14}$ could not be
constrained without separate knowledge of the equation of state.
However, in the example just mentioned one could serendipitously
tightly constrain both the equation of state and the value of
$c_{14}$, since a redshift of 0.35 is the lower limit of all the
curves at this mass.

\section{Discussion}
\label{conclusions}

We have found that the structure of non-rotating neutron stars in
Einstein-aether theory is fairly close to that in general
relativity, but there are quantitative differences. Depending on
the equation of state, the maximum masses range from about 6-15\%
smaller than in GR when the ae-theory parameter $c_{14}$ is equal
to 1. The corresponding surface redshifts are roughly 10\% larger
than in GR. Measurements of high gravitational masses or precise
surface redshifts have the potential to yield strong joint
constraints on the equation of state and on deviations from GR.
Therefore, as laboratory experiments and other observations narrow
down the equation of state of cold matter at several times nuclear
density, neutron star observations may be a valuable resource for
exploring deviations from general relativity in strong gravity.

We now make some comments about related further work that would be
interesting to pursue. First, the neutron star solutions we
considered are at rest with respect to the asymptotic aether.
Although corrections due to motion with respect to the aether are
not significant for the present paper, they are important for the
high precision predictions of radiation damping in compact binaries.
In particular, the missing ingredient in the analysis of
\cite{Foster:2007gr} is the value of the ``sensitivity" parameter
measuring the velocity dependence of the mass. It should be possible
to compute this parameter for different masses and different
equations of state by determining the velocity perturbations of the
solutions found here, or by finding the exact nonlinear solutions
with finite velocity.

We considered in this paper only neutron star phenomenology. What
is the situation for black holes? It was found in
\cite{Eling:2006ec} that although different from stellar
exteriors, nonrotating black hole solutions in ae-theory are very
similar to the Schwarzschild solution of GR. Hence it is not
likely that these could lead to significant constraints.

Nevertheless, it is  quite conceivable that strong deviations from
GR will be found for rapidly rotating black hole solutions. This
is suggested by the presence of the ergoregion, in which the
inertial frames are strongly dragged. The preferred frame aspects
of ae-theory may be conspicuous here, due to larger gradients in
the aether field. Also, unlike for the spherically symmetric
non-rotating case, the spin-1 degrees of freedom of the aether
could be activated in the axially symmetric setting. To explore
these issues would require finding numerical solutions describing
rotating black holes in Einstein-aether theory.

\section*{Acknowledgments} We are grateful to Sudip Bhattacharyya for
kindly providing us with tables characterizing the nuclear
equations of state used in our analysis. This work was supported
in part by the National Science Foundation under grants
PHY-0601800 and AST-0607428.


\begin{thebibliography}{99}

\bibitem{Kramer06}
M. Kramer et al., ``Tests of General Relativity from Timing the
Double Pulsar," Science {\bf 314}, 97 (2006).

\bibitem{Iwasawa96}
K. Iwasawa et al., ``The variable iron K emission line in
MCG-6-30-151996," Mon. Not. R. Astron. Soc. {\bf 282}, 1038 (1996).

\bibitem{Fab02}
A.~C. Fabian et al., ``A long hard look at MCG-6-30-15 with
XMM-Newton," Mon. Not. R. Astron. Soc. {\bf 335}, L1 (2002).

\bibitem{RN03}
C.~S. Reynolds and M.~A. Nowak, ``Fluorescent iron lines as
a probe of astrophysical black hole systems," Phys. Rep.
{\bf 377}, 389 (2003).

\bibitem{RBG05}
C.~S. Reynolds, L.~W. Brenneman and D. Garofalo, ``Black Hole Spin
in AGN and GBHCs," Astrophys. Space Sci. {\bf 300}, 71 (2005).

\bibitem{BR06}
L.~W. Brenneman and C.~S. Reynolds, ``Constraining Black Hole Spin
via X-Ray Spectroscopy," Astrophys. J. {\bf 652}, 1028 (2006).

\bibitem{White06}
N. White, ``The Constellation-X Mission," High Resolution X-ray
Spectroscopy: towards XEUS and Con-X, Proceedings of the
international workshop held at the Mullard Space Science Laboratory
of University College London, Holmbury St Mary, Dorking, Surrey, UK,
March 27 - 28, 2006, Ed. G. Branduardi-Raymont, E41 (2006)
http://www.mssl.ucl.ac.uk/$\sim$gbr/workshop2/.

\bibitem{Barish00}
B.~C. Barish, ``The Laser Interferometer Gravitational-Wave
Observatory LIGO," Adv. Space Res. {\bf 25}, 1165 (2000).

\bibitem{Fidecaro97}
F. Fidecaro et al., ``The Virgo Interferometer for Gravitational
Wave Detection," in Proc. 12th Italian Conf. on General Relativity
and Gravitational Physics, ed. M. Bassan, V. Ferrari, M.
Francaviglia, F. Fucito and I. Modena (Singapore: World Scientific),
163 (1997).

\bibitem{Schilling98}
R. Schilling, ``The GEO600 Ground-Based Interferometer for
the Detection of Gravitational Waves," in AIP Conf. Proc. 456,
Laser Interferometer
Space Antenna, Second International LISA Symp., ed.
W. M. Folkner (New York: AIP), 217 (1998).

\bibitem{Ando02}
M. Ando et al., ``Current status of TAMA," Class. Quant. Grav.
{\bf 19}, 1409 (2002).

\bibitem{Danzmann00}
K. Danzmann, ``LISA Mission Overview," Adv. Space Res. {\bf 25}, 1129 (2000).

\bibitem{scalartensor}
A.~Salmona, Phys.\ Rev.\ D {\bf 154} 1218 (1967); W.~Hillebrant and
H.~Heintzmann, Gen.\ Relativ.\ Gravitation {\bf 5} 663 (1974)

%\cite{Damour:1993hw}
\bibitem{Damour:1993hw}
  T.~Damour and G.~Esposito-Farese,
  ``Nonperturbative strong field effects in tensor - scalar theories of
  gravitation,''
  Phys.\ Rev.\ Lett.\  {\bf 70}, 2220 (1993).

  %\cite{DeDeo:2003ju}
\bibitem{DeDeo:2003ju}
  S.~DeDeo and D.~Psaltis,
  ``Towards New Tests of Strong-field Gravity with Measurements of Surface
  Atomic Line Redshifts from Neutron Stars,''
  Phys.\ Rev.\ Lett.\  {\bf 90}, 141101 (2003)
  [arXiv:astro-ph/0302095].

%\cite{Clayton:1999zs}
\bibitem{Clayton:1999zs}
  M.~A.~Clayton and J.~W.~Moffat,
  ``Scalar-Tensor Gravity Theory For Dynamical Light Velocity,''
  Phys.\ Lett.\ B {\bf 477}, 269 (2000)
  [arXiv:gr-qc/9910112].
  %%CITATION = GR-QC 9910112;%%

%\cite{Jacobson:2000xp}
\bibitem{Jacobson:2000xp}
  T.~Jacobson and D.~Mattingly,
  ``Gravity with a dynamical preferred frame,''
  Phys.\ Rev.\ D {\bf 64}, 024028 (2001)
  [arXiv:gr-qc/0007031].
  %%CITATION = GR-QC 0007031;%%

 %\cite{Arkani-Hamed:2003uy}
\bibitem{Arkani-Hamed:2003uy}
  N.~Arkani-Hamed, H.~C.~Cheng, M.~A.~Luty and S.~Mukohyama,
  ``Ghost condensation and a consistent infrared modification of gravity,''
  JHEP {\bf 0405}, 074 (2004)
  [arXiv:hep-th/0312099].
  %%CITATION = HEP-TH 0312099;%%

 %\cite{Gripaios:2004ms}
\bibitem{Gripaios:2004ms}
  B.~M.~Gripaios,
  ``Modified gravity via spontaneous symmetry breaking,''
  JHEP {\bf 0410}, 069 (2004)
  [arXiv:hep-th/0408127].
  %%CITATION = HEP-TH 0408127;%%
  %%Cited 23 times in SPIRES-HEP

 %\cite{Bluhm:2004ep}
\bibitem{Bluhm:2004ep}
  R.~Bluhm and V.~A.~Kostelecky,
  ``Spontaneous Lorentz violation, Nambu-Goldstone modes, and gravity,''
  Phys.\ Rev.\ D {\bf 71}, 065008 (2005)
  [arXiv:hep-th/0412320].
  %%CITATION = HEP-TH 0412320;%%

%\cite{Heinicke:2005bp}
\bibitem{Heinicke:2005bp}
  C.~Heinicke, P.~Baekler and F.~W.~Hehl,
  ``Einstein-aether theory, violation of Lorentz invariance, and  metric-affine
  gravity,''
  Phys.\ Rev.\ D {\bf 72}, 025012 (2005)
  [arXiv:gr-qc/0504005].
  %%CITATION = GR-QC 0504005;%%

%\cite{Rubakov:2006pn}
\bibitem{Rubakov:2006pn}
   V.~A.~Rubakov,
  ``Phantom without UV pathology,''
  Theor.\ Math.\ Phys.\  {\bf 149}, 1651 (2006)
  [Teor.\ Mat.\ Fiz.\  {\bf 149}, 409 (2006)]
  [arXiv:hep-th/0604153].


  %\cite{Cheng:2006us}
\bibitem{Cheng:2006us}
  H.~C.~Cheng, M.~A.~Luty, S.~Mukohyama and J.~Thaler,
  ``Spontaneous Lorentz breaking at high energies,''
  JHEP {\bf 0605}, 076 (2006)
  [arXiv:hep-th/0603010].
  %%CITATION = HEP-TH 0603010;%%

%\cite{Eling:2004dk}
\bibitem{Eling:2004dk}
  C.~Eling, T.~Jacobson and D.~Mattingly,
  ``Einstein-aether theory,''
 in {\sl Deserfest}, eds. J.~Liu, M.~J.~Duff, K.~Stelle, and
R.~P.~Woodard (World Scientific, 2006)
  arXiv:gr-qc/0410001.
  %%CITATION = GR-QC 0410001;%%

%\cite{Carroll:2004ai}
\bibitem{Carroll:2004ai}
S.~M.~Carroll and E.~A.~Lim, ``Lorentz-violating vector fields
slow the universe down,'' Phys.\ Rev.\ D {\bf 70}, 123525 (2004)
[arXiv:hep-th/0407149].
%%CITATION = HEP-TH 0407149;%%

  %\cite{Eling:2003rd}
\bibitem{Eling:2003rd}
  C.~Eling and T.~Jacobson,
  ``Static post-Newtonian equivalence of GR and gravity with a dynamical
  preferred frame,''
  Phys.\ Rev.\ D {\bf 69}, 064005 (2004)
  [arXiv:gr-qc/0310044].
  %%CITATION = GR-QC 0310044;%%

%\cite{Graesser:2005bg}
\bibitem{Graesser:2005bg}
  M.~L.~Graesser, A.~Jenkins and M.~B.~Wise,
  ``Spontaneous Lorentz violation and the long-range gravitational
  preferred-frame effect,''
  Phys.\ Lett.\ B {\bf 613}, 5 (2005)
  [arXiv:hep-th/0501223].
  %%CITATION = HEP-TH 0501223;%%

%\cite{Foster:2005dk}
\bibitem{Foster:2005dk}
  B.~Z.~Foster and T.~Jacobson,
  ``Post-Newtonian parameters and constraints on Einstein-aether theory,''
  Phys.\ Rev.\ D {\bf 73}, 064015 (2006)
  [arXiv:gr-qc/0509083].
  %%CITATION = GR-QC 0509083;%%

%\cite{Jacobson:2004ts}
\bibitem{Jacobson:2004ts}
T.~Jacobson and D.~Mattingly, ``Einstein--Aether waves,'' Phys.\
Rev.\ D {\bf 70}, 024003 (2004) [arXiv:gr-qc/0402005].
%%CITATION = GR-QC 0402005;%%

%\cite{Lim:2004js}
\bibitem{Lim:2004js}
  E.~A.~Lim,
  ``Can We See Lorentz-Violating Vector Fields in the CMB?,''
  Phys.\ Rev.\ D {\bf 71}, 063504 (2005)
  [arXiv:astro-ph/0407437].
  %%CITATION = ASTRO-PH 0407437;%%

%\cite{Elliott:2005va}
\bibitem{Elliott:2005va}
  J.~W.~Elliott, G.~D.~Moore and H.~Stoica,
  ``Constraining the new aether: Gravitational Cherenkov radiation,''
  JHEP {\bf 0508}, 066 (2005)
  [arXiv:hep-ph/0505211].
  %%CITATION = HEP-PH 0505211;%%

%\cite{Eling:2005zq}
\bibitem{Eling:2005zq}
  C.~Eling,
  ``Energy in the Einstein-aether theory,''
  Phys.\ Rev.\ D {\bf 73}, 084026 (2006)
  [arXiv:gr-qc/0507059].
  %%CITATION = GR-QC 0507059;%%

%\cite{Foster:2006az}
\bibitem{Foster:2006az}
  B.~Z.~Foster,
  ``Radiation damping in Einstein-aether theory,''
  Phys.\ Rev.\ D {\bf 73}, 104012 (2006)
  [arXiv:gr-qc/0602004].
  %%CITATION = GR-QC 0602004;%%

%\cite{Foster:2007gr}
\bibitem{Foster:2007gr}
  B.~Z.~Foster,
  ``Strong field effects on binary systems in Einstein-aether theory,''
  arXiv:0706.0704 [gr-qc].
  %%CITATION = ARXIV:0706.0704;%%

%\cite{Eling:2006df}
\bibitem{Eling:2006df}
  C.~Eling and T.~Jacobson,
  ``Spherical Solutions in Einstein--Aether Theory: Static Aether and Stars,''
  Class.\ Quant.\ Grav.\  {\bf 23}, 5625 (2006)
  [arXiv:gr-qc/0603058].
  %%CITATION = GR-QC 0603058;%%

 %\cite{Eling:2006ec}
\bibitem{Eling:2006ec}
  C.~Eling and T.~Jacobson,
  ``Black holes in Einstein-aether theory,''
  Class.\ Quant.\ Grav.\  {\bf 23}, 5643 (2006)
  [arXiv:gr-qc/0604088].
  %%CITATION = GR-QC 0604088;%%

%\cite{Cook:1993qr}
\bibitem{Cook:1993qr}
  G.~B.~Cook, S.~L.~Shapiro and S.~A.~Teukolsky,
  ``Rapidly rotating neutron stars in general relativity: Realistic equations
  of state,''
  Astrophys.\ J.\  {\bf 424}, 823 (1994).
  %%CITATION = ASJOA,424,823;%%

%\cite{Akmal:1998cf}
\bibitem{Akmal:1998cf}
  A.~Akmal, V.~R.~Pandharipande and D.~G.~Ravenhall,
  ``The equation of state for nucleon matter and neutron star structure,''
  Phys.\ Rev.\ C {\bf 58}, 1804 (1998)
  [arXiv:nucl-th/9804027].
  %%CITATION = NUCL-TH 9804027;%%

%\cite{Farhi:1984qu}
\bibitem{Farhi:1984qu}
  E.~Farhi and R.~L.~Jaffe,
  ``Strange Matter,''
  Phys.\ Rev.\  D {\bf 30}, 2379 (1984).
  %%CITATION = PHRVA,D30,2379;%%

\bibitem{Bhattprivate} Sudip Bhattacharyya,
  private communication.

\bibitem{Quaintrell03}
H. Quaintrell et al., ``The mass of the neutron star in Vela X-1 and
tidally induced non-radial oscillations in GP Vel," Astron.
Astrophys. {\bf 401}, 313 (2003).

\bibitem{Nice05}
D.~J. Nice et al., ``A $2.1\,M_\odot$ Pulsar Measured by
Relativistic Orbital Decay," Astrophys. J. {\bf 634}, 1242 (2005).

\bibitem{Barret05a}
D. Barret, J.-F. Olive and M.~C. Miller, ``Drop of coherence of the
lower kilo-Hz QPO in neutron stars: Is there a link with the
innermost stable circular orbit?," Astron. Nachr. {\bf 326}, 808
(2005).

\bibitem{Barret05b}
D. Barret, J.-F. Olive and M.~C. Miller, ``An abrupt drop in the
coherence of the lower kHz quasi-periodic oscillations in 4U
1636-536," Mon. Not. R. Astron. Soc. {\bf 361}, 855 (2005).

\bibitem{Barret06}
D. Barret, J.-F. Olive and M.~C. Miller, ``The coherence of
kilohertz quasi-periodic oscillations in the X-rays from accreting
neutron stars," Mon. Not. R. Astron. Soc.  {\bf 370}, 1140 (2006).

\bibitem{Barret07}
D. Barret, J.-F. Olive and M.~C. Miller, ``Supporting evidence for
the signature of the innermost stable circular orbit in Rossi X-ray
data from 4U 1636-536," Mon. Not. R. Astron. Soc. {\bf 376}, 1139
(2007).

\bibitem{Mendez06}
M. M\'endez, ``On the maximum amplitude and coherence of the
kilohertz quasi-periodic oscillations in low-mass X-ray binaries,"
Mon. Not. R. Astron. Soc. {\bf 371}, 1925 (2006).

\bibitem{Cottam02}
J. Cottam, F. Paerels and M. M\'endez, ``Gravitationally redshifted
absorption lines in the X-ray burst spectra of a neutron star,"
Nature (London) {\bf 420}, 51 (2002).

\bibitem{Ozel:2006km}
F.~\"Ozel, ``Soft equations of state for neutron-star matter ruled
out by EXO 0748-676," Nature (London) {\bf 441}, 1115 (2006).

\end{thebibliography}
\end{document}